\documentclass[epjST]{svjour}
\usepackage{graphics}
\usepackage{xcolor}
\usepackage{amsmath}
\usepackage{hyperref}
\usepackage{amssymb}
\usepackage{soul}

\newcommand{\gsim}{\raisebox{-0.13cm}{~\shortstack{$>$ \\[-0.07cm]
      $\sim$}}~}
\newcommand{\lsim}{\raisebox{-0.13cm}{~\shortstack{$<$ \\[-0.07cm]
      $\sim$}}~}

\begin{document}
\title{Modeling and characterizing stochastic neurons based on in vitro voltage-dependent spike probability functions}
\author{Vinicius Lima\inst{1,2}\fnmsep\thanks{\email{vinicius.lima.cordeiro@gmail.com}} \and Rodrigo F. O. Pena\inst{1,3}\and Renan O. Shimoura\inst{1} \and Nilton L. Kamiji\inst{1} \and Cesar C. Ceballos\inst{1,4}\and Fernando  S. Borges\inst{5,7} \and Guilherme S. V. Higa\inst{5} \and  Roberto de Pasquale\inst{6} \and Antonio C. Roque\inst{1}\fnmsep\thanks{\email{antonior@usp.br}}}
\institute{Department of Physics, Faculty of Philosophy, Sciences and Letters of Ribeir\~{a}o Preto, University of S\~{a}o Paulo, Ribeir\~{a}o Preto, SP, Brazil \and Institut de Neurosciences des Systèmes (INS, UMR 1106) Université de Aix-Marseille, Marseille, France \and Federated Department of Biological Sciences, New Jersey Institute of Technology and Rutgers University, Newark, New Jersey, NJ, USA \and Vollum Institute, Oregon Health \& Science University, Portland, OR, USA \and Center for Mathematics, Computation, and Cognition, Federal University of ABC, S\~{a}o Bernardo do Campo, SP, Brazil \and Department of Physiology and Biophysics, Institute of Biomedical Sciences, University of S\~{a}o Paulo, S\~{a}o Paulo, SP, Brazil \and State University of New York Downstate Health Sciences University, NY, USA}

\abstract{
Neurons in the nervous system are submitted to distinct sources of noise, such as ionic-channel and synaptic noise, which introduces variability in their responses to repeated presentations of identical stimuli. This motivates the use of stochastic models to describe neuronal behavior. In this work, we characterize an intrinsically stochastic neuron model based on a voltage-dependent spike probability function. We determine the effect of the intrinsic noise in single neurons by measuring the spike time reliability and study the stochastic resonance phenomenon. The model was able to show increased reliability for non-zero intrinsic noise values, according to what is known from the literature,  and the addition of intrinsic stochasticity in it enhanced the region in which stochastic-resonance is present.  We proceeded to the study at the network level where we investigated the behavior of a random network composed of stochastic neurons. In this case, the addition of an extra dimension, represented by the intrinsic noise, revealed dynamic states of the system that could not be found otherwise. Finally, we propose a method to estimate the spike probability curve from in vitro electrophysiological data. }
%
\maketitle
\section{Introduction}
\label{intro}

Modeling the nervous system is challenging
not only due to the size of the brain \cite{herculano2015mammalian} but also because of the complex patterns of neuronal connectivity across the different brain regions \cite{hagmann2008mapping}. Such intricate structure is successful in processing diverse types of information: motor activities, face recognition, cognition, etc. However, the complex degree of organization gives rise to a susceptibility to several sources of noise \cite{faisal2008}, which range from channel noise \cite{movshon2000reliability,white2000channel,GirardiSchappow2013} to synaptic noise \cite{destexhe2012,pena2018b} and network noise \cite{DesPar99,LonRot10}. Noise may also lead to a decrease in the accuracy of brain connectivity inference \cite{nunes2019}. It is therefore important to model neurons taking into account such stochastic internal and external sources to study their effects on information processing. 

The influence of noise on neurons is well documented experimentally. \textit{In vitro} electrophysiological recordings show that when a given neuron is stimulated many times by the same input,
its response is not the same across trials. There is variability in the spike times \cite{faisal2008,brette2003reliability,ermentrout2008reliability}, though the mean spike count over the stimulation time is approximately constant. One of the sources of this variability is the intrinsic stochasticity present in the ion channels at the neuronal membrane \cite{faisal2008,white2000channel}. The  effects of the intrinsic noise can be observed as subthreshold voltage fluctuations \cite{steinmetz2000subthreshold} that might affect action potential emission timing \cite{jacobson2005subthreshold,protachevicz2020}, threshold variability \cite{sigworth1980variance,azouz1999cellular}, and the probability of spontaneous spikes \cite{chow1996spontaneous}. It is possible to reduce the neuronal spike variability by increasing the number of ionic channels in the membrane \cite{white2000channel}, but there is always a trade-off between the number of ion channels, ion pumps, and their metabolic cost \cite{laughlin1998metabolic,attwell2001energy}. 

How the brain deals with the presence of noise to produce reliable processing is not known. A possible explanation is that the minimization of the influence of noise would be done by averaging over neuronal populations. According to this mechanism, neurons would harness the redundancy present in the input signal to represent it using a population coding that when averaged would reduce the influences of individual sources of noise \cite{faisal2008}. 
It was shown that adding a fluctuating signal to a DC input current increases the reliability of spike times across trials (the fluctuating signal must be the same across trials, hence it is called ``frozen signal'') \cite{mainen1995reliability,stevens1998input}. This could be a way whereby a given neuronal population would respond reliably to oscillatory inputs coming from other brain regions \cite{buracas1998efficient}.

Much is argued about the functional role of noise in the brain \cite{faisal2008,stein2005neuronal,mcdonnell2011benefits}. 
Experiments show that channel noise may be crucial to: (i) determine reliability in spike times \cite{schneidman1998ion}; (ii) influence the dynamics of entorhinal cortex neurons \cite{white1998noise}; (iii) cause changes in firing patterns of sensory neurons \cite{braun1994oscillation,braun1997low}; and (iv) perithreshold oscillations in entorhinal stellate neurons \cite{dorval2005channel}.
The fact that the brain seems to use noise in certain functions might be due to its evolution under the influence of noise \cite{faisal2008,mcdonnell2009stochastic}. An interesting example in this direction is the phenomenon of stochastic resonance (SR), where non-linear systems can optimally enhance the detection of low-amplitude oscillatory inputs for a certain level of noise \cite{mcdonnell2009stochastic,dykman1998can}. This phenomenon has been observed both experimentally \cite{douglas2004neuronal,ward2010stochastic}, and theoretically \cite{stocks2001generic,schmerl2013channel}

The introduction of noise in models can be done in different ways. A standard approach is to consider a deterministic model, which obeys a set of differential equations, and add stochastic processes as input sources. 
Alternatively, spikes can be randomly generated by modeling the neuron as an intrinsically stochastic element. These ways of modeling may be equivalent in particular situations such as when a deterministic system has a stochastic threshold, mimicking the so called escape rate model \cite{gerstner2014,PleGer00}, or by introducing voltage-dependent spike probability functions \cite{galves2013,brochini2016}. Although such functions are usually theoretically selected, they can be determined based on electrophysiological recordings \cite{jolivet2006predicting}. 
In the present work, we describe an intrinsically stochastic neuron model that takes into account functional neuronal mechanisms such as stochastic resonance. We also show how this model can be used as a tool to study the relationship between stochasticity at the single neuron level and network behavior. Additionally, we propose a method by which the firing probability function can be estimated from electrophysiological recordings.

This paper is organized as follows: In the methods section we introduce the stochastic neuron model using an exponential voltage-dependent spike probability function \cite{jolivet2006predicting}. Then, in the results section we study the behavior of the single neuron focusing on what phenomena the intrinsic noise can reproduce. Next, we use these neurons to implement a stochastic version of a well-know network with random architecture \cite{brunel2000dynamics}. We evaluate how the intrinsic noise changes the network behavior as a function of its parameters, namely the frequency of the background input and the relative strength of inhibitory synapses. Finally, we suggest a method to extract the spike probability function from electrophysiological data. We finish the paper by discussing the main results observed and possible applications of our description.

\section{Methods}
\label{sec:methods}

\subsection{Single neuron model description}\label{neuron}

The subthreshold dynamics of the neuron model follows the ``leaky integrate-and-fire'' (LIF) formalism, and is given by Eq.~\ref{eqn:lif_din}.

\begin{equation}
\frac{dV(t)}{dt} = -\frac{V(t) - V_{\rm r}}{\tau_{\text{m}}} + \frac{I_{\text{inj}}(t)}{C_\text{m}} 
\label{eqn:lif_din}
\end{equation}  

\noindent where $V$ is the membrane potential, $C_{\rm m}$  is the membrane capacitance, $V_{\rm r}$ is the resting membrane potential, $I_{\rm inj}$ is an external current injected into the neuron, and $\tau_{\text{m}}$ is the membrane time constant.

To implement an action potential, the LIF model defines a threshold voltage $V_{\rm th}$ and sets a reset rule so that every time $V$ crosses $V_{\rm th}$ it is considered that the neuron fired. In the model that used here, the emission of a spike is stochastic and depends on a voltage-dependent firing intensity \cite{PerGer67,gerstner2014} $\phi(V)$, which will be called here ``spike probability function'', defined by Eq.~\ref{eqn:exp_phi}.

\begin{equation}
	\phi(V) = \frac{1}{b} \exp{\left[\frac{V-V_{1/2}}{a}\right]},
	\label{eqn:exp_phi}
\end{equation}

\noindent where $V_{1/2}$ is the voltage at which  $\phi(V)=1/b$, and $b$ and $a$ are parameters \cite{jolivet2006predicting}. After emitting a spike the membrane potential is held at the reset value, $V_{\rm{reset}}$, for the absolute refractory time $\tau_{\rm ref}$. Notice that the stochasticity level is controlled by the parameter $a$, which will be called ``intrinsic stochasticity'' parameter.

\subsection{Network model}\label{network}

The network model used in the present work follows the random connectivity scheme of the classic model of Brunel \cite{brunel2000dynamics}: $N$ neurons are divided into two populations where $N_{\rm e} = 0.8N$ are excitatory neurons and $N_{\rm i} = 0.2N$ are inhibitory neurons. Each neuron receives $C = 0.1 N$ connections chosen randomly with the imposition that $C_{\rm e} = 0.8C$ are excitatory and $C_{\rm i} = 0.2C$ are inhibitory. 

Apart from the internal connections of the network, each neuron receives $C_{\rm e}^{\rm ext} = C_{\rm e}$ synaptic inputs from an external, not explicitly modeled, population, with the same synaptic weight (equal to the internal connection weight $J$, see below). These inputs are modeled as independent Poisson processes with homogeneous rate $\nu_{\rm ext}$. 

In this model the synaptic input from the pre-synaptic neuron $j$ to the post-synaptic neuron $i$ is given by Eq.~\ref{eqn:sin_brunel}.

\begin{equation}
\frac{I_\text{i, \text{syn}}}{C_\text{m}} = \sum_{j} J_{ij}\sum_{k}\delta(t-t_{j}^{f,k}-d),
\label{eqn:sin_brunel}
\end{equation}

\noindent where $J_{ij}$ is the synaptic weight of the $j \rightarrow i$ connection, $t_{j}^{f, k}$ is the time of the $k^{\rm th}$ spike from neuron $j$, and $d$ is the transmission delay. All excitatory connections have the same weight $J$. For the inhibitory connections the weight is multiplied by the factor $-g$, called the relative strength of inhibitory connections. All fixed parameters are shown in Table~\ref{tab:tab1}.

\begin{table}[!h]
\centering
\caption{Parameters of the stochastic neuron model (Eqs.~\ref{eqn:lif_din}-\ref{eqn:exp_phi}) and of the random network.}
\label{tab:tab1}   
\begin{tabular}{|l|l|l|l|l|l|l|l|l|l|l|}
	\hline
			$V_\text{th}$ & $V_\text{r}$ &  $V_\text{reset}$ & $\tau_\text{m}$ &  $\tau_\text{ref}$& $N_\text{e}$ & $N_\text{i}$ & $C_\text{e}$ & $C_\text{i}$ & $J$ & $d$ \\
			\hline 
		$20$ mV  & $0$ mV & $10$ mV & $20$ ms & $2$ ms & $10000$ & $2500$ & $1000$ & $250$ & $0.1$ mV & $2$ ms \\
		\hline 
\end{tabular}
\end{table}

All codes used to implement the neuron and network models were written in Python. The implementation of neuron and network models were done using the
neurosimulator Brian 2 \cite{goodman2009}. The codes are available at:\\ \href{github.com/ViniciusLima94/stochastic\_neuron\_model}{github.com/ViniciusLima94/stochastic\_neuron\_model}.

\subsection{Spike-train statistics}\label{measures}

The spike train $x(t)$ of a given neuron is defined as $x(t) = \sum_{\{t^f\}}\delta(t - t^f)$, where $\{t^f\}$ is the set of spike times of the neuron and $\delta(t)$ is the Dirac delta function. 

The mean spike-count (firing rate) over a time interval $T$ can be obtained from the spike-train as  $f = (1/T)\int_{T}x(t)dt = N_{\text{spikes}}/T$.

The irregularity of a spike-train is measured with the coefficient of variation ($\text{CV}$) of the interspike intervals (ISIs), defined by $\text{CV} = \sigma_{\langle \text{ISI} \rangle}/\langle \text{ISI} \rangle$, where $\sigma_{\langle \text{ISI} \rangle}$ is the standard deviation and $\langle \text{ISI} \rangle$ is the mean of the distribution of ISIs. The neuron is considered to fire irregularly if $\text{CV} \gsim 1$, and regularly if  $\text{CV} \ll 1$. 

To quantify the degree of synchrony among neurons in a network, we use the phase locking value ($\text{PLV}$) which is a standard measure to evaluate phase synchronization \cite{lachaux1999,celka2007,rosenblum2011,aydore2013,lowet2016}. We define the $\text{PLV}$ as the average over $K$ neuron pairs and $T$ sample time points: 
\begin{equation}
\text{PLV} = \frac{1}{K} \sum_{\{ij\}}^K\left|\sum_{t}^T e^{i \Delta \Phi_{xy}(t)} \right|,   
\label{Eq:sync_index_plv}
\end{equation} 
where $\Delta \Phi_{xy}(t)$ are the phase differences $\Phi_{x}(t)-\Phi_{y}(t)$ from two randomly chosen spike-trains $\left(x(t),y(t)\right)$ that are obtained using the Hilbert transform. The $\text{PLV}$ is bounded between 0 (asynchronous) and 1 (synchronous). 

To quantify whether there is oscillatory activity or not in the network we compute the spectral entropy $H_{S}$, defined by Eq.~ \ref{eq:spec_ent}.

\begin{equation}
	H_{S} = -\frac{\sum_{k=0}^{m} S_{xx}(f)\ln(S_{xx}(f))}{\ln{N_{f}}}
	\label{eq:spec_ent}
\end{equation}

\noindent where $S_{xx}(f)$ is the averaged spike-train power spectrum computed via Fourier transform and $N_f$ is the number of  frequencies in the power spectrum. $H_{s}$ is constrained between 0 (all the power concentrated in a single frequency) and 1 (flat broadband spectrum).

The correlation coefficient between the spike-train $x(t)$ and a given input signal $S(t)$ is computed as $CC = \text{cov}(x, S)/\sqrt{\text{var}(x)\text{var}(S)}$ \cite{gabbiani1998principles}. 

\subsection{Electrophysiological Recordings}\label{recordings}

The following refers to the experimental data used to fit the spike probability function. Whole-cell patch clamp recordings were conduct using male wistar rats with $20-25$ postnatal days. All animal were kept at 12:12 h light-dark cycle in an  animal house (Institute of Biomedical Sciences - University of S\~ao Paulo) with temperature adjusted for $23^{o}{\rm C}\pm2^{o}$ with free access to food and water.  All procedures were accepted by the Institutional Animal Care Committee of the Institute of Biomedical Sciences, University of S\~ao Paulo (CEUA ICB/USP n. 090, fls. $1^{o}$).

Once animals were anesthetized by means of isoflurane inhalation (AErrane; Baxter Pharmaceuticals) they were decapitated and the brain was quickly removed and subsequently submerged in cooled ($0-2^{o}$C) oxygenated (5\% CO2-95\% O2) cutting solution (in mM): 206 sucrose, 25 NaHCO3, 2.5 KCl, 10 MgSO4, 1.25 NaH2PO4, 0.5 CaCl2, and 11 D-glucose). Before slices preparation, cerebellum and brain hemispheres were removed.  Both brain hemispheres were than glued in a metal platform and sectioned using a vibratome (Leica- VT1200). 350-400$\mu$m brain slices were obtained by advancing the vibratome blade (0.08mm/s, amplitude 0.95mm) from anterior-posterior orientation. Slices were rapidly transferred to a chamber containing an oxygenated (5\% CO2-95\% O2) artificial cerebrospinal fluid (ACSF; in mM): 125 NaCl, 25 NaHCO3, 3 KCl, 1.25 NaH2PO4, 1 MgCl2, 2 CaCl2, and 25 D-glucose). Slices were kept oxygenated at room temperature (20-25$^{o}$C) for at least one hour before proceeding with electrophysiological recordings.

Brain slices containing the hippocampal formation were placed in a submersion-type recording chamber upon a modified microscope stage. All procedure was realized with a maintained constant perfusion of oxygenated ACSF (5\% CO$_2$-95\% O$_2$) at 30$^{o}$C. Patch-clamp whole-cell recordings were made from pyramidal-shape neurons positioned in CA1 pyramidal layer. Recording pipettes were manufactured from borosilicate glass (Garner Glass) with input resistances of $\approx$4-6 M$\Omega$.  Pipettes were filled with intracellular solution (in mM): 135 K-gluconate, 7 NaCl, 10 HEPES, 2 Na2ATP, 0.3 Na3GTP, 2 MgCl2; at a pH of 7.3 adjusted with KOH and osmolality of 290 mOsm. All experiments were performed using a visualized slice setup under a differential interference contrast-equipped Nikon Eclipse E600FN microscope. Recordings were made by using a Multiclamp 700B amplifier and pClamp software (Axon Instruments). Only recordings from cells that presented spontaneous activity with membrane potentials lower than -60 mV, access resistance lower than 20 M$\Omega$, and input resistance higher than 100 M$\Omega$ and lower than 1000 M$\Omega$ were analyzed. In order to identify different spiking patterns (regular, tonic, or bursting spike) cells were injected with depolarizing currents. Neuronal spontaneous activity was assessed by 10 minutes of continuous recordings in current-clamp mode.

\section{Results}

\subsection{Single neuron}

In this section we will focus on the single neuron level analysis of the stochastic model, primarily aiming to verify if the model reproduces well known phenomena. Particularly, we reproduce the phenomenon of stochastic resonance due to intrinsic noise. In this analysis we use $b = 27$ in Eq.~\ref{eqn:exp_phi}.

\begin{figure}[!h]
\centering
\resizebox{1\columnwidth}{!}{\includegraphics{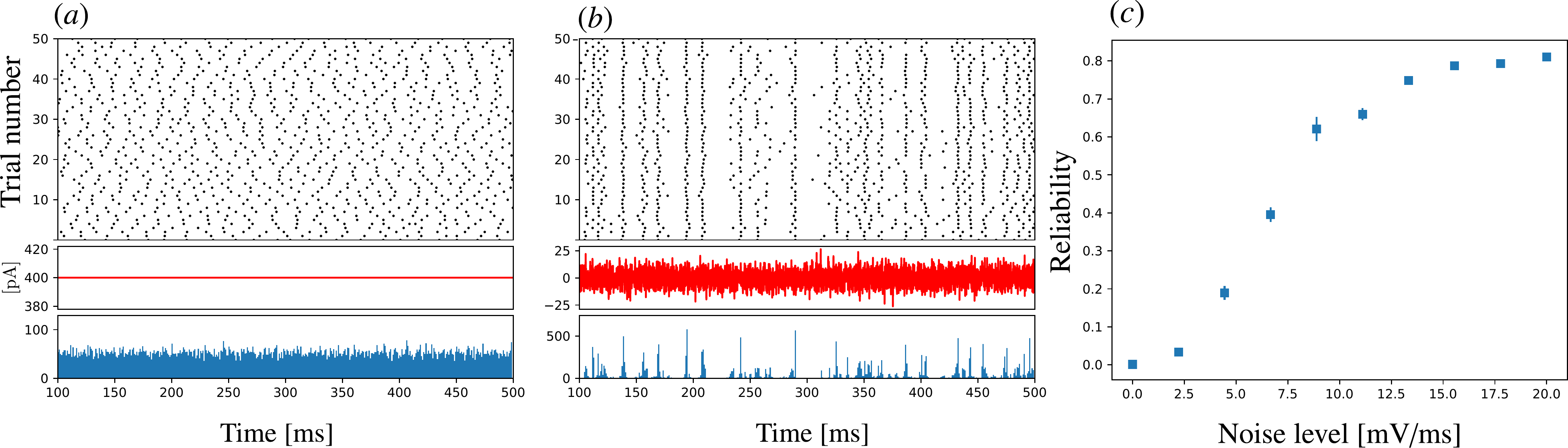} }
\caption{Computing the reliability of the single neuron model. (a) Raster plot (top) and PSTH (bottom) for 50 trials (chosen out of 10000) in which the simulated neuron received the same constant DC input current of 400 pA as stimulus (middle). In the raster plot, time is shown in the horizontal axis and trial number along the vertical axis. The dots indicate spike times. The PSTH is the histogram of the number of spikes for time intervals of length $\Delta t =$ 1 ms divided by the number of trials and $\Delta t$. (b) Same as in (a) but now the neuron received a fluctuating input $I_{\text{noise}}$ (see text) with noise level $\eta$. We used $\eta = 7$ in the plot. (c) Plot of the reliability $R$ of the stochastic neuron receiving a constant DC current of 400 pA plus an external noise input $_{\text{noise}}$ as a function of the noise level $\eta$. The reliability is defined as the sum of the number of spikes in each bin over all bins divided by the total number of spikes possible (at most one spike can occur per bin per trial). We do this only for bins where the activity is greater than $70\%$ of the highest rate in the PSTH.}
\label{fig:raster_single}
\end{figure}

The stochastic behavior of the neuron model can be seen by inspecting the raster plot, which shows spike times for successive presentations of the same stimulus. Fig.~\ref{fig:raster_single}(a) shows the raster plot when the stimulus is a constant DC current $I_{\rm DC} = 400$ pA. One can see that the spike times across trials are highly variable, and the peristimulus time histogram (PSTH) at the bottom is 
``flat'' indicating that the spike times are not reliably reproduced over trials.

Next, we add an external noise input $I_{\text{noise}}$ to the DC current. This input is given by $I_{\text{noise}} = \eta\xi(t)$, where $\xi(t)$ is a Gaussian white noise process with zero mean and standard deviation $\eta$ (noise level) in mV/ms. The raster plot when the noise level is $\eta = 7$ mV/ms is shown in Fig.~\ref{fig:raster_single}(b). The corresponding PSTH (bottom panel) shows several peaks indicating that the spike times tend to be more repeatable across trials. To quantify the reliability of spike times as a function of the noise level, we use the ``reliability'' measure $R$ introduced by Mainen and Sejnowski~\cite{mainen1995reliability}, which is defined as the fraction of the maximum number of possible spikes in periods of high firing rates. Fig.~\ref{fig:raster_single}(c) shows the reliability measure as a function of $\eta$. $R$ grows monotonically towards 1 as $\eta$ increases, replicating, at least qualitatively, the behavior observed experimentally by Mainen and Sejnowski (see Fig. 2C of \cite{mainen1995reliability}). This differs from a deterministic model, e.g. the LIF model, where for $\eta$ sufficiently high the reliability would be one, and zero otherwise.

Our next step is a study of stochastic resonance (SR). SR is a phenomenon in which the detection of oscillatory subthreshold signals in a non-linear system is enhanced by the presence of an optimal level of noise. Usually, SR is detected by plotting the signal-to-noise ratio (SNR) of a signal as a function of the noise level. Systems that exhibit SR show a maximum in the SNR for a noise level greater than zero \cite{mcdonnell2009stochastic}. Here we are interested in testing whether the intrinsically stochastic neuron model is capable of manifesting SR. Experimental \cite{BezVod95} and computational \cite{schmerl2013channel} works show that channel noise is capable of generating SR. To do so, we vary the intrinsic stochasticity parameter $a$ in Eq.~\ref{eqn:exp_phi} and, to measure the resonance, instead of the SNR we compute the correlation coefficient (CC) between the neuron spike train for each trial and a sinusoidal subthreshold input with frequency of $10$ Hz and amplitude $I_{\rm a} = 300$ pA as a function of $a$.

\begin{figure}[!h]
\centering
\resizebox{1\columnwidth}{!}{\includegraphics{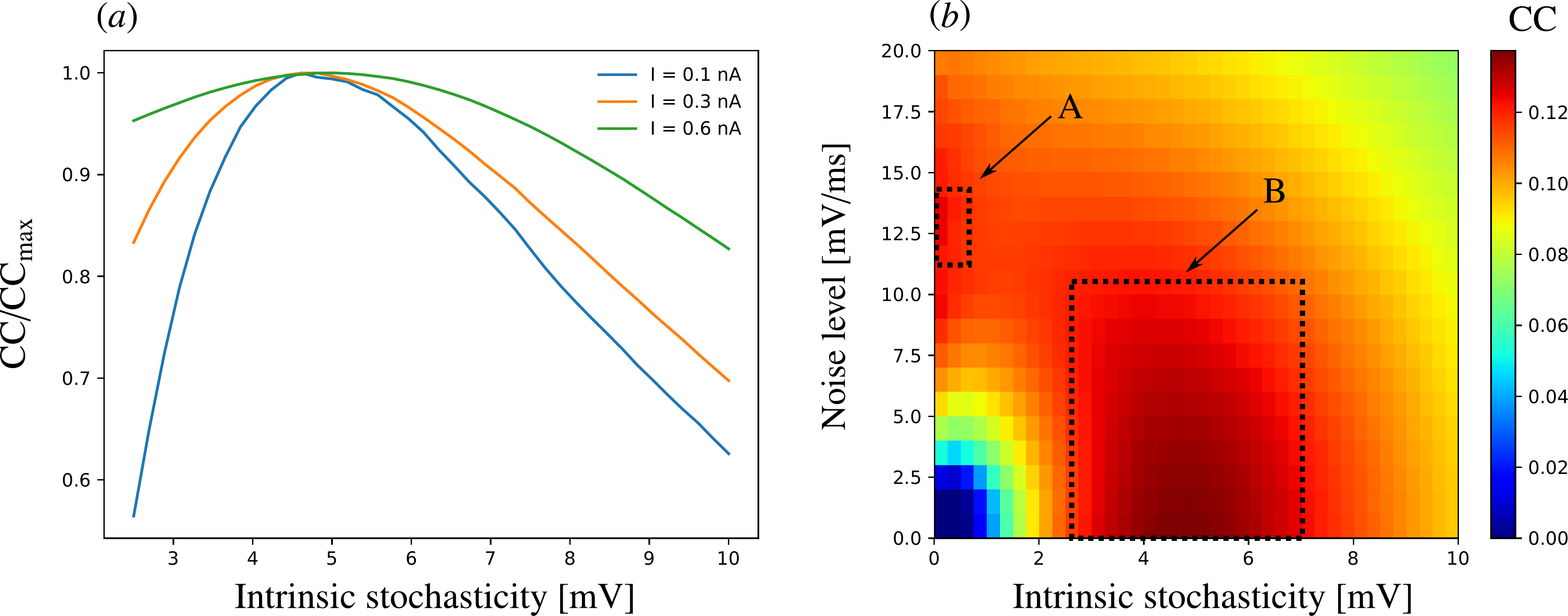} }
\caption{Stochastic resonance in the single neuron model. (a) Correlation coefficient (normalized by its maximum) between the neuron spike train and a sinusoidal input current of frequency 10 Hz and amplitude $I$ as a function of the intrinsic stochasticity parameter $a$ (see Eq.~\ref{eqn:exp_phi}). Three different values of $I$ were used (shown on the upper right corner of the plot): 0.1 (blue), 0.3 (orange) and 0.6 nA (green); the first two correspond to sub-threshold stimuli and the latter to a supra-threshold stimulus. (b) map of the correlation coefficient as a function of the intrinsic stochasticity $a$ and the noise level $\eta$ of the fluctuating signal added to the sinusoidal input. The dashed rectangles indicate regions A and B (see text).}
\label{fig:sr}
\end{figure}

In Fig.~\ref{fig:sr}(a), we show the CC (normalized by its maximum) as a function of $a$ for three different values of $I_{\rm a}$. For subthreshold amplitudes ($I_{\rm a} = 100$ and $300$ pA) the neuronal response displays SR, visualized as a maximum in the CC for $a \approx 5$ mV. On the other hand, for a suprathreshold amplitude ($600$ pA) the resonance peak vanishes, indicating that the inclusion of noise does not act as an agent to facilitate signal detection in this scenario. The reason for the latter effect is easily explained by the fact that the higher the current the higher the spike frequency over all values of $a$. Resonance is the amplification of a signal in an intermediate stochastic value $a$. For that, the signal-to-noise ratio can not be high. For stronger currents it is expected that the signal will achieve a state where firing is found over all values of $a$ in a way that there will be no amplification when noise is present (high signal-to-noise ratio). In this particular case for $600$ pA, the current is already above the rheobase (firing at high frequencies) and the noise only adds to the input current, thus not acting as a spike-facilitator mechanism (such as in SR).

Further, we compare the effects of obtaining SR via an extrinsic noise term $I_{\text{noise}}$ added to the sinusoidal input and via the intrinsic noise of the neuron model. In Fig.~\ref{fig:sr}(b) we show the CC as a function of two parameters: the noise level $\eta$ of the external noise input and the intrinsic stochasticity level $a$. For a deterministic neuron model ($a = 0$), the maximum of CC occurs for $12 \leq \eta \leq 13$ (see region A in Fig.~\ref{fig:sr}(b)). Interestingly, for intrinsically stochastic neuron models ($a > 0$) the region where external noise causes SR is amplified (see region B in Fig.~\ref{fig:sr}(b)). This suggests that SR becomes more robust as a function of the joint effect of the intrinsic and extrinsic stochasticity sources.

\begin{figure}[!h]
\centering
\resizebox{1\columnwidth}{!}{\includegraphics{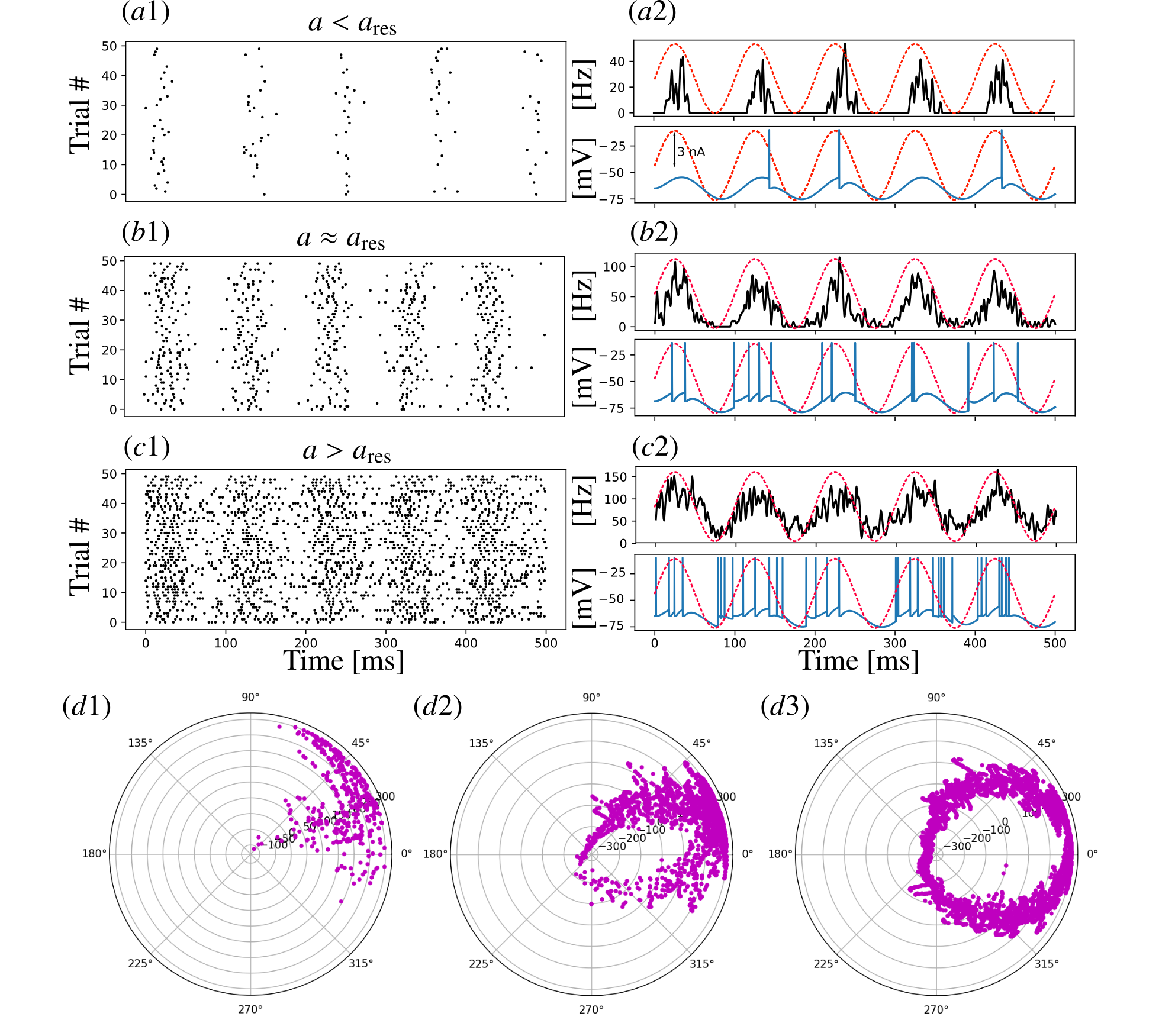} }
\caption{Stochastic resonance at an ``optimal'' level of intrinsic sochasticity. All plots in this figure correspond to the stochastic neuron model submitted to a sinusoidal input current with frequency 10 Hz and amplitude $I_{\rm a} = 300$ pA. (a1, b1 and c1) raster plots for three values of the intrinsic stochasticity level $a$: (a1) $a = 2$ mV $< a_{\rm res}$, (b1) $a = 5$ mV $= a_{\rm res}$, and (c1) $a = 10$ mV $> a_{\rm res}$. (a2, b2, and c2) PSTH for the corresponding raster plots in (a1, b1, c1) (top panels) and the membrane potential of a randomly selected neuron (bottom panel). The red dashed line is the sinusoidal input. (d1, d2, d3) polar scatter plots in which each dot is placed as a function of the sinusoidal input amplitude $I_{\rm a}(t)$, and the relative phase ($\Delta\Phi$) between the PSTH and the input signal at the spike times, for the PSTHs in (a2, b2 and c2) respectively. In the direction from center to periphery, the concentric circles in d1 run from $-100$ pA to 300 pA, and in d2 and d3 they run from $-300$ pA to 300 pA.}
\label{fig:raster_pot_sr}
\end{figure}

Results of an extended characterization of the behavior of the stochastic neuron model in the presence of a noiseless oscillatory input signal are shown in Fig.~\ref{fig:raster_pot_sr}. Raster plots and PSTHs are shown in Figs.~\ref{fig:raster_pot_sr}(a1,b1,c1) and Figs.~\ref{fig:raster_pot_sr}(a2,b2,c2) respectively. For a value of the intrinsic stochasticity parameter $a$ below resonance ($a = 2$ mV), the raster plot displays a low frequency periodic firing with few spikes during the peak phases of the oscillatory input. The corresponding PSTH activity is above zero only when the sinusoidal current is near its maximum amplitude. The membrane potential of a neuron chosen randomly from the population displays subthreshold oscillations following the sinusoidal input with a few occasional spikes at a frequency of approximately $0.6$ spikes per cycle. At resonance ($a = 5$ mV), the periodicity seen in the raster plot and PSTH is maintained but the frequency is higher and the number of spikes at the peaks of the oscillatory input is larger. The PSTH activity follows the oscillatory time course of the input more reliably and the chosen neuron emits spikes at a higher frequency of approximately $2.6$ spikes per cycle. Above resonance ($a = 10$ mV), a periodicity is still visually discernible in the raster plot, but the spikes are more evenly distributed over time. For example, at the troughs of the oscillatory input the PSTH activity is relatively high ($\approx 40$ Hz) and the single neuron emits many spikes.

A study of the relationship between the spike times and the amplitude and phase of the sinusoidal input is shown in Figs.~\ref{fig:raster_pot_sr}(d1,d2,d3). The polar scatter plots show well defined regions corresponding to the three situations: below resonance, at resonance and above resonance. Below resonance, (Fig.~\ref{fig:raster_pot_sr}(d1)), spikes occur mostly for $I(t) > 100$ pA and the relative phase between the PSTH and the sinusoidal input signal is almost completely restricted to the quadrant between $0^{\circ}$ and $90^{\circ}$. At resonance, spikes can occur for both low and high amplitudes of the input signal and are distributed over a ``pinch-like'' region of the scatter plot (Fig.~\ref{fig:raster_pot_sr}(d2)). The density of spikes is higher at the 
``upper branch'' of the pinch, which runs mostly along the line of $45^{\circ}$. Above resonance, the pinch assumes a ring-like shape and spikes are homogeneously scattered along it.  

\subsection{Network}

In this section, we study the behavior of the network described in section \ref{network} with stochastic neurons as described in section \ref{neuron}. The objective is to characterize how stochasticity at the single neuron level influences the dynamics of the network. To quantify the dynamics of the network we use the measures defined in section \ref{measures} as a function of the network parameters, namely the frequency of the Poisson external input ($\nu_{\rm ext}$), and the relative strength of inhibitory synapses ($g$). 

\begin{figure}[!h]
\centering
\resizebox{1\columnwidth}{!}{\includegraphics{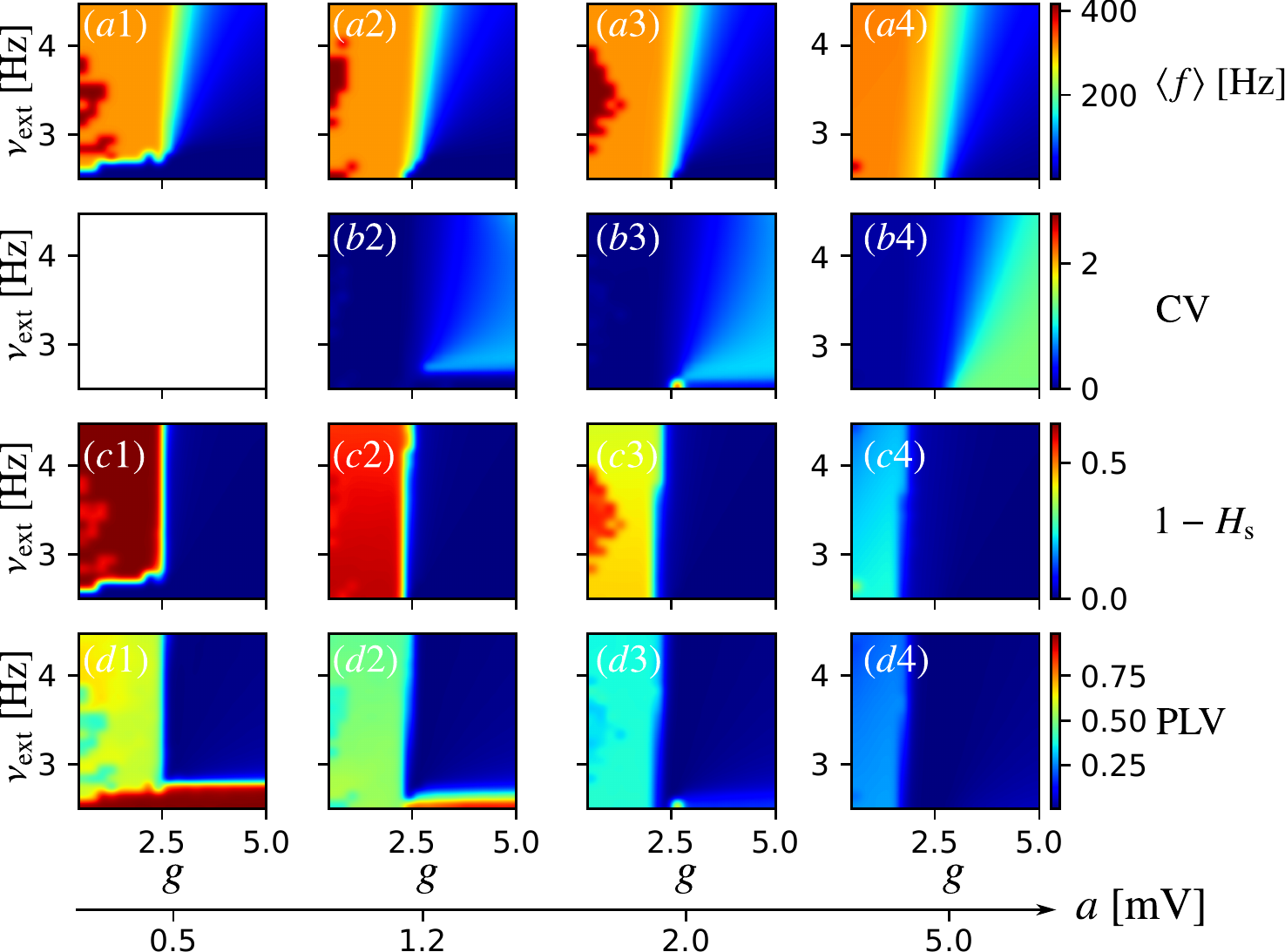} }
\caption{Phase diagrams of the network for different values of the intrinsic stochasticity parameter $a$. In all diagrams the horizontal axis indicates the relative inhibitory synaptic strength $g$ and the vertical axis indicates the frequency of the external Poisson input ($\nu_{\rm ext}$). The columns correspond to increasing values of $a$ as indicated in the axis at the bottom. From top to bottom rows, the diagrams give the average network firing rate (a1-a4), the average CV (b1-b4), $1 - H_s$ (c1-c4), and average PLV (d1-d4). The values of these measures are given by a color code shown at the vertical bars to the right of the diagrams.}
\label{fig:network_maps}
\end{figure}

In Figs.~\ref{fig:network_maps}(a1--a4) we show the average firing rate of the network for increasing values of the stochasticity level $a$. 
At $g \approx 2.5$, the network changes from a regime of high firing rate ($\langle f \rangle > 200$ Hz) to a regime where the firing rates are lower than $200$ Hz. For low $g$ ($\lsim 1.3$), the network can have firing rates close to $400$ Hz, and the size of the region where these high rates occur seems to reach a maximum at $a = 2.0$ mV and then decreases for higher $a$. For the nearly deterministic neuron ($a = 0.5$ mV), at low values of $\nu_{\rm ext}$ the firing rates are closer to zero. As the stochasticity parameter $a$ increases, the network can be active even for low frequency inputs. 

In Figs.~\ref{fig:network_maps}(b1--b4) we show the diagrams of average CV (or irregularity). For $a = 0.5$ mV, there is a black stripe at the bottom of the diagram corresponding to the region where the network has low firing rate. Since most neurons do not fire in this region the CV cannot be computed. In general, the irregularity increases as $g$ crosses the transition point $g \approx 2.5$, and it becomes more prominent as $a$ increases. Notice that the greater $\nu_{\rm ext}$, the greater $g$ must be for the irregularity to be above 1. This is due the fact that for regions with high firing rates the neurons tend to spike at every time step, therefore the network spiking becomes regular.

The oscillatory activity of the network (Figs.~\ref{fig:network_maps}(c1--c4)), is related to the firing rate behavior in Figs.~\ref{fig:network_maps}(a1--a4). In general, higher frequencies imply more well defined oscillations. For frequencies higher than $200$ Hz, neurons tend to spike regularly and more synchronously as shown in the diagrams in Figs.~\ref{fig:network_maps}(d1--d4). The degree of synchrony is high for both high and low frequencies. This happens because if every neuron spikes or is silent most of the time the collective behavior of the network is synchronous. Interestingly, for $a = 2.0$ mV the CV and PLV diagrams display a single point at $\nu_{\rm ext} = 0$ Hz, and $g \approx 2.5$ where the network activity is irregular and synchronous. 

\subsection{Method for estimating the parameters of the voltage-dependent firing probability function}\label{sec::method_fit_phi}

The algorithm developed to determine the voltage-dependent spike probability function $\phi(V)$ uses long time series. The electrophysiological data used consisted of $10$ minutes of spontaneous activity of a CA1 pyramidal neuron (see section \ref{recordings}), making it suitable for our goals. 

\begin{figure}[!h]
\centering
\resizebox{.8\columnwidth}{!}{\includegraphics{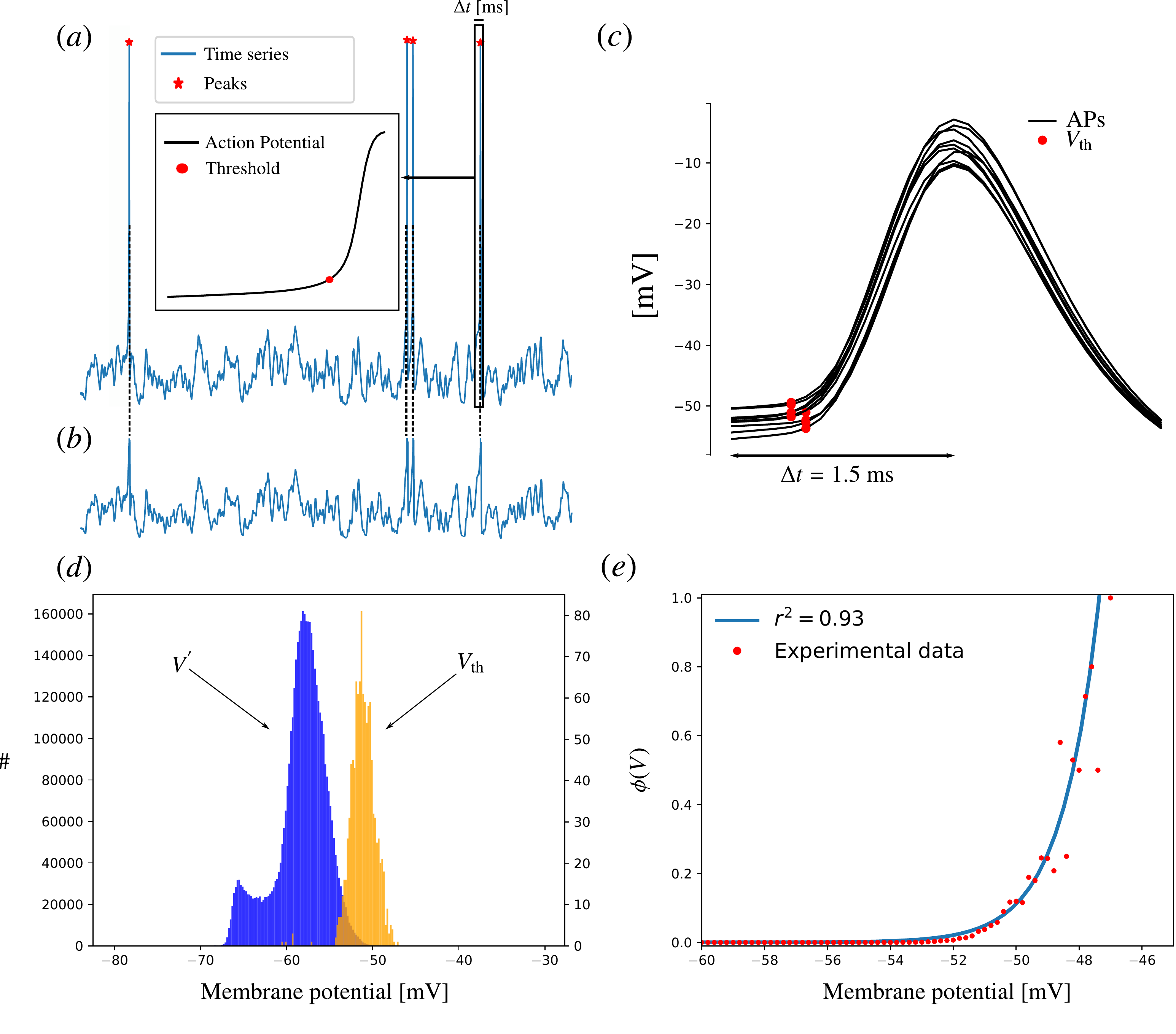} }
\caption{Schematic representation of the method used to determine the firing probability function. (a) Detection of the threshold values $V_{th}$. Given a voltage time series, the peaks are detected and a time window of length $\Delta t$ is placed around each peak. For each time window the $V_{th}$ is determined using Equation~\ref{eqn:curv} (see inset). (b) Time series after the removal of the membrane potential values greater than the corresponding thresholds. (c) The plot shows action potentials randomly selected from the time series with the time window $\Delta t$ used to determine the thresholds, which are indicated by red dots. (d) Histograms of the clipped membrane potentials ($V^{'}(t)$ (in blue) and thresholds $V_{\rm th}(t)$ (in red). (e) Estimated $\phi(V)$ from the experimental data (red dots) and the exponential function (blue curve) used to fit the points.}
\label{fig:ex_algoritmo}
\end{figure}

The first step of the algorithm is to determine the values of the membrane potential time series $V(t)$ at the onsets of the action potentials, i.e. the threshold values ($V_{\rm th}$) \cite{platkiewicz2010threshold}. There are several different methods for estimating $V_{\rm th}$, and the one adopted here is considered one of the best \cite{sekerli2004estimating}. According to this method, the action potential threshold corresponds to the point of maximum curvature ($K_{\rm p}$) of the time series $V(t)$ in the moments preceding the emission of an action potential. The curvature of a function $V(t)$ is defined by 

\begin{equation}
K_\text{p} = \ddot{V}[1 + \dot{V}^{2}]^{-\frac{3}{2}},
\label{eqn:curv}
\end{equation}

\noindent where $\dot{V}$ and $\ddot{V}$ indicate the first and second derivatives of the time series $V(t)$, respectively. The threshold is defined as:

\begin{equation}
	V_{\rm th} = \operatorname*{argmax}_{V} K_{\rm p}
	\label{eqn:thr_def}
\end{equation} 

To determine the points of maximum curvature in the time series $V(t)$, first we detect all the peaks above $-10$ mV in $V(t)$, which correspond to the action potentials. Then, for each peak $i$ we place a time window $\Delta t$ with its rightmost tip exactly on the peak time (so that it covers a time interval shortly before the action potential). After that we compute the maximum curvature of the time series within the time window using Eqs.~\ref{eqn:curv} and~\ref{eqn:thr_def}. At the end we have a list containing all thresholds for the time series analyzed. In Fig.~\ref{fig:ex_algoritmo}(a) we illustrate the procedure described above.

After determining all $V_{\rm thr}$ values, we remove the ascending and descending parts of the action potentials from the time series $V(t)$ by clipping the time series at the heights given by $V_{\rm thr}$. This results in a second time series $V^{'}(t)$ as illustrated in Fig.~\ref{fig:ex_algoritmo}(b).

Finally, two histograms are built: one for the membrane potentials in the series $V^{'}(t)$, and another one for $V_{\rm th}$. Both histograms are superposed as in Fig. 4(d), and the ring intensity is estimated as the binwise ratio $V_{\rm thr}/V^{'}(t)$ between the two histograms. We fitted the resulting points by the exponential function $\phi(V)$ in Eq.~\ref{eqn:exp_phi} (see Fig.~\ref{fig:ex_algoritmo}(e)). The parameters $a$, $b$, and  $V_{1/2}$ of the exponential function obtained with the fitting are, respectively, $1.19$ mV, $27.0$ mV, and $-51.3$ mV.

\section{Discussion}

Modeling neurons can be challenging due to the large number of variables one needs to obtain. The task is particularly difficult due to the stochastic nature of neuronal spiking~\cite{faisal2008,destexhe2012,white2000channel}. In this work, we systematically characterized a simple intrinsically stochastic neuron model with a reduced number of variables over different case scenarios. We also showed how to calibrate such a model from {\it in vitro} experiments.

Our observation of the behavior of an isolated neuron was focused on the voltage-dependent probability function which defines its behavior. In terms of the concept of reliability $R$, we showed that the model can reproduce experimentally reported results \cite{mainen1995reliability} when stimulated by a noisy input. 

Moreover, we found that the model can display stochastic resonance. Our exploration of such case was based on considering the intrinsic noise of the neuron as a variable parameter. In addition, we have tested the possibility of achieving stochastic resonance by the joint effects of the intrinsic stochasticity of the neuron model combined with the extrinsic stochasticity provided by an external noisy input. This increased the parameter space of detection of the neuron so that it became more sensitive to subthreshold inputs. This result suggests a possible way by which neurons can use their intrinsically generated noise plus external noisy inputs to enhance signal detection. 

We also provided an example result in which the stochastic resonance observed in the neuron model happens due to the fact that the number of spikes emitted near the maximum of a sinusoidal input is maximized giving information about both the period and shape of the input signal. Moreover, we showed that there is a relation between the phase of the peristimulus time histogram and the sinusoidal signal dependent on how close (or distant) the intrinsic noise level of the neuron is from its resonance value.

In terms of population behavior, we have analysed a random network of excitatory and inhibitory neurons described by the stochastic neuron model. We focused on the dynamics of the network by quantifying its average frequency, irregularity, synchrony and oscillatory behavior as a function of the network parameters and the intrinsic stochasticity of the neuron model. Our results showed that the intrinsic stochasticity of the neuron can affect the dynamic behavior of the network. It is already known that intrinsic firing properties of single neurons can have an impact on network properties~\cite{tomov2014sustained}, and our results here suggest that the added firing variability due to the intrinsic noise sources can also produce observable effects. 

Finally, we provided a method to fit the voltage-dependent spike probability function of stochastic neuron from time series of voltage recordings. The method was used to adjust the parameters of the stochastic neuron model from {\it in vitro} data. Methodological uncertainties are expected when measuring the voltage-dependent spike probability experimentally. From the case studied, we noticed that even more data than the ones considered here would be needed to estimate the firing probability for $V > -50$ mV. In the example considered here, we used the maximum curvature of the voltage time series to determine the action potential onset but other criteria could be used as well.

In conclusion, the different studies presented here show that the intrinsically stochastic neuron model of Eqs.~\ref{eqn:lif_din} and~\ref{eqn:exp_phi} is a viable alternative to deterministic LIF models for implementing spiking neural network models. The model captures a number of phenomena associated to intrinsic sources of neuronal noise and can be used in studies on the impact of single neuron spike stochasticity on network dynamic behavior. Future works in this field could improve the model by (i) including richer subthreshold dynamics to the model (e.g., by including a second ODE for the spike-adaptation variable), in order to allow it to have a wider repertoire of behaviors; (ii) adding other relevant variables to determine the spike probability function, such as, the first derivative of the membrane potential and the inter-spike intervals. 

\section*{Acknowledgements}

This article was produced as part of the IRTG 1740/TRP 2011/50151-0, funded by the DFG/FAPESP. It was also supported partially by the S. Paulo Research Foundation
(FAPESP) Research, Innovation and Dissemination Center
for Neuromathematics (CEPID NeuroMat, Grant No. 2013/07699-0). 
The authors also thank FAPESP support through Grants Nos. 2013/25667-8 (R.F.O.P.),
2015/50122-0 (A.C.R.),
2016/03855-5 (N.L.K.), 
2017/07688-9 (R.O.S), 
2017/18977-1 (F.S.B),  
2018/20277-0 (A.C.R.), and
2019/14962-5 (R.P).
V.L. and C.C.C. were supported by a CAPES PhD scholarship.
A.C.R. thanks financial support from the National Council of Scientific
and Technological Development (CNPq), Grant No. 306251/2014-0. This study was financed in part by the Coordenação de Aperfeiçoamento de Pessoal de Nível Superior - Brasil (CAPES) - Finance Code 001.

\section*{Author contribution statement}

Author Contributions: V.L., R.F.O.P., R.O.S.,   N.L.K., C.C.C., and A.C.R.: Conceived the work; V.L., R.F.O.P., R.O.S., C.C.C, and N.L.K.: work on model implementation, simulation, and theoretical analysis V.L., R.F.O.P., R.O.S.,  N.L.K., C.C.C., and A.C.R.: wrote the manuscript; F.S.B., G.S.V.H., R.P., collected and provided the data used in the study. All authors read, reviewed, and agreed to the published version of the manuscript.

%


\begin{thebibliography}{}
	\bibitem{herculano2015mammalian}
	S. Herculano-Houzel, K. Catania, P.R. Manger, and J.H. Kaas, Brain Behav. Evol. \textbf{86}, (2015) 145--163. 
	
	\bibitem{hagmann2008mapping}
	P. Hagmann, L. Cammoun, X. Gigandet, R. Meuli, C.J. Honey, V.J. Wedeen, and O. Sporns, PLoS Biol. \textbf{6}, (2008) e159. 
	
	\bibitem{faisal2008}
	A.A. Faisal, L.P.J. Selen, and D.M. Wolpert, Nat. Rev. Neurosci. \textbf{9}, (2008) 292.
	
	\bibitem{movshon2000reliability}
	J.A. Movshon, Neuron \textbf{27}, (2000) 412--414.
	
	\bibitem{white2000channel}
	J.A. White, J.T. Rubinstein, and A.R. Kay,  Trends Neurosci. \textbf{23}, (2000) 131--137.
	
	\bibitem{GirardiSchappow2013}
	M. Girardi-Schappo, O. Kinouchi, and M.H.R. Tragtenberg, Phys. Rev. E \textbf{88}, (2013) 024701
	
	\bibitem{destexhe2012}
	A. Destexhe, and M. Rudolph-Lilith, \textit{Neuronal Noise} (Springer, 2012).
	
	\bibitem{pena2018b}
	R.F.O. Pena, M.A. Zaks, and A.C. Roque, J. Comput. Neurosci. \textbf{45}, (2018) 1--28.
	
	\bibitem{DesPar99}
	A. Destexhe, and D. Paré, J. Neurophysiol. \textbf{81}, (1999) 1531--1547.
	
	\bibitem{LonRot10}
	M. London, A. Roth, L. Beeren, M. H\"{a}usser, and P.E. Latham, Nature. \textbf{466}, (2010) 123--127.
	
	\bibitem{nunes2019}
	R.V. Nunes, M.B. Reyes, and R.Y. de Camargo, Biol Cybern \textbf{113}, (2019) 309–320
	
	\bibitem{brette2003reliability}
	R.Brette, and E. Guigon, Neural Comput. \textbf{15}, (2003) 279--308.
	
	\bibitem{ermentrout2008reliability}
	G.B. Ermentrout, R.F. Galán, and N.N. Urban, Trends Neurosci. \textbf{31}, (2008) 428--434.
	
	\bibitem{steinmetz2000subthreshold}
	P.N. Steinmetz, A. Manwani, C. Koch, M. London, and I. Segev,  J. Comput. Neurosci. \textbf{9}, (2000) 133--148.
	
	\bibitem{jacobson2005subthreshold}
	G.A. Jacobson, K. Diba, A. Yaron-Jakoubovitch, Y. Oz, C. Koch, I. Segev, and Y. Yarom, J. Physiol. \textbf{564}, (2005) 145--160.
	
	\bibitem{protachevicz2020}
	P.R. Protachevicz, M.S. Santos, E.G. Seifert, E.C. Gabrick, F.S. Borges, R.R. Borges, J. Trobia, J.D. Szezech Jr, K.C. Iarosz, I.L. Caldas, C.G. Antonopoulos, Y. Xu, R.L. Viana, A.M. Batista, arXiv preprint, (2020) arXiv:2005.14597
	
	\bibitem{sigworth1980variance}
	F.J. Sigworth, J. Physiol. \textbf{307}, (1980) 97--129.
	
	\bibitem{azouz1999cellular}
	R. Azous, and C. M. Gray, J. Neurosci. \textbf{19}, (1999) 2209--2223.
	
	\bibitem{chow1996spontaneous}
	C.C. Chow, and J.A. White, Biophys. J. \textbf{71}, (1996) 3013--3021.
	
	\bibitem{laughlin1998metabolic}
	S.B. Laughlin, R.R.R. van Steveninck, and J.C. Anderson, Nat. Neurosci. \textbf{1}, (1998) 36--41.
	
	\bibitem{attwell2001energy}
	D. Attwell, and S.B. Laughlin, J Cerebr. Blood F. Met. \textbf{21}, (2001) 1133--1145.
	
	\bibitem{mainen1995reliability}
	Z.F. Mainen, and T.J. Sejnowski, Science \textbf{268}, (1995) 1503--1506.
	
	\bibitem{stevens1998input}
	C.F. Stevens, and A.M. Zador. Nat. Neurosci. \textbf{1}, (1998) 210--217.
	
	\bibitem{buracas1998efficient}
	G.T. Buracas, A.M. Zador, M.R. DeWeese, and T.D. Albright, Neuron \textbf{20}, (1998) 959--969.
	
	\bibitem{stein2005neuronal}
	R.B. Stein, E.R. Gossen, and K.E. Jones, Nat. Rev. Neurosci. \textbf{6}, (2005) 389--397.
	
	\bibitem{mcdonnell2011benefits}
	M.D. McDonnel, and L.M. Ward, Nat. Rev. Neurosci. \textbf{12}, (2011) 415--425.
	
	\bibitem{schneidman1998ion}
	E. Schneidman, B. Freedman, and I. Segev, Neural Comput. \textbf{10}, (1998) 1679--1703.
	
	\bibitem{white1998noise}
	J.A. White, R. Klink, A. Alonso, and A.R. Kay, J. Neurophysiol. \textbf{80}, (1998) 262--269.
	
	\bibitem{braun1994oscillation}
	H.A. Braun, H. Wissing, K. Sch\"{a}fer, and M.C. Hirsch, Nature \textbf{367}, (1994) 270--273.
	
	\bibitem{braun1997low}
	H.A.Braun, K. Sch\"{a}fer, K. Voigt, R. Peters, F. Bretschneider, X. Pei, L. Wilkens, and F. Moss, J. Comput. Neurosci.\textbf{4}, (1997) 335--347.
	
	\bibitem{dorval2005channel}
	A.D. Dorval, and J.A. White, J. Neurosci. \textbf{25}, (2005) 10025--10028.
	
	\bibitem{mcdonnell2009stochastic}
	M.D. McDonnell, and D. Abbott, PLoS Comput. Biol. \textbf{5}, (2009) e1000348.
	
	\bibitem{dykman1998can}
	M.I. Dykman, and P.V.E.McClintock, Nature \textbf{391}, (1998) 344.
	
	\bibitem{douglas2004neuronal}
	R.J. Douglas, and K.A.C. Martin, Annu. Rev. Neurosci. \textbf{27}, (2004) 419--451.
	
	\bibitem{ward2010stochastic}
	L.M. Ward, S.E. MacLean, and A. Kirschner, PLoS one \textbf{5}, (2010) e14371.
	
	\bibitem{stocks2001generic}
	N.G. Stocks, and R. Mannella, Phys. Rev. E \textbf{64}, (2001) 030902.
	
	\bibitem{schmerl2013channel}
	B.A. Schmerl, and M.D. McDOnnell, Phys. Rev. E \textbf{88}, (2013) 052722.
	
	\bibitem{gerstner2014}
	W. Gerstner, and W.M. Kistler, \textit{Neuronal dynamics: From single neurons to networks and models of cognition} (Cambridge University Press, 2014).
	
	\bibitem{PleGer00}
	H.E. Plesser, and W.Gerstner, Neurocomp. \textbf{32}, (2000) 219--224.
	
	\bibitem{galves2013}
	A. Galves, and E. L\"{o}cherbach, J. Stat. Phys. \textbf{151}, (2013) 896--921.
	
	\bibitem{brochini2016}
	L. Brochini, A.A. Costa, M. Abadi, A.C. Roque, J. Stolfi, and O.Kinouchi, Sci. Rep. \textbf{6}, (2016) 35831.
	
	\bibitem{jolivet2006predicting}
	R. Jolivet, A. Rauch, H.R. L\"{u}scher, and W. Gerstner, J.Comput. Neurosci. \textbf{21}, (2006) 35--49.
	
	\bibitem{brunel2000dynamics}
	N. Brunel, J. Comput. Neurosci. \textbf{8}, (2000) 183--208.
	
	\bibitem{PerGer67}
	D.H. Perkel, G.L. Gerstein, and G.P. Moore, Biophys. \textbf{7}, (1967) 391--418.
	
	\bibitem{goodman2009}
	D.F.M. Goodman, and R. Brette, Front. Neurosci., \textbf{3}, (2009) 26.
	
	\bibitem{lachaux1999}
	J.P. Lachaux, E. Rodriguez, J. Martinerie, and F.J. Varela, Hum. Brain. Mapp. \textbf{8}, (1999) 194--208.
	
	\bibitem{celka2007}
	P. Celka, IEEE Signal Proc. Let. \textbf{14}, (2007) 577--580.
	
	\bibitem{rosenblum2011}
	M. Rosenblum, A. Pikovsky, J. Kurths, C. Sch\"{a}fer, and P.A. Tass, \textit{Handbook of biological physics} (Elsevier, 2001) 279--321. 
	
	\bibitem{aydore2013}
	S. Aydore, D. Pantazis, and R. M. Leahy, NeuroImage \textbf{74}, (2013) 231--244.
	
	\bibitem{lowet2016}
	E. Lowet, M.J. Roberts, P. Bonizzi, J. Karel, and P.D. Weerd, PLoS one \textbf{11}, (2016) e0146443.
	
	\bibitem{gabbiani1998principles}
	F. Gabbiani, and C. Koch, Methods in neuronal modeling \textbf{12}, (1998) 313--360.
	
	\bibitem{BezVod95}
	S.M. Bezrukov, and I. Vodyanoy, Nature \textbf{378}, (1995) 362--364.
	
	\bibitem{platkiewicz2010threshold}
	J. Platkiewicz, and R. Brette, PLoS Comput. Biol. \textbf{6}, (2010) e1000850.
	
	\bibitem{sekerli2004estimating}
	M. Sekerli, C.A. Del Negro, R. H. Lee, and R.J. Butera, IEEE T. Bio-Med Eng. \textbf{51}, (2004) 1665--1672.
	
	\bibitem{tomov2014sustained}
	P. Tomov, R.F.O. Pena, M.A. Zaks, and A.C. Roque, Front. Comput. Neurosci. \textbf{8}, (2014) 103.
	
		
	
\end{thebibliography}


\end{document}